\documentclass[prd,nofootinbib,preprintnumbers,
amsmath,amssymb,aps,
reprint,
]{revtex4-1}

\usepackage{graphicx}
\usepackage{dcolumn}
\usepackage{bm}
\usepackage{color}
\input{colordvi.tex}

\newcommand{\beq}{\begin{equation}}
\newcommand{\eeq}{\end{equation}}
\newcommand{\beqa}{\begin{eqnarray}}
\newcommand{\eeqa}{\end{eqnarray}}

\begin{document}

\preprint{APS/123-QED}

\title{Cosmological constraints on the velocity-dependent baryon-dark matter coupling}

\author{Junpei Ooba${}^1$}
\email{ooba.jiyunpei@f.mbox.nagoya-u.ac.jp}
\author{Horoyuki Tashiro${}^1$}
\author{Kenji Kadota${}^2$}

\affiliation{
${}^1$Department of physics and astrophysics, Nagoya University, Nagoya 464-8602, Japan\\ 
${}^2$Center for Theoretical Physics of the Universe, Institute for Basic Science (IBS), Daejeon 34051, Korea
}

\date{\today}

\begin{abstract}
We present the cosmological constraints on the cross section of baryon-dark matter interactions
for the dark matter mass below the MeV scale from the Planck CMB (cosmic microwave background)
and SDSS (Sloan Digital Sky Survey) Lyman-$\alpha$ forest data. 
To explore the dark matter mass $m_{\chi}\lesssim 1$ MeV
for which the dark matter's free-streaming effect can suppress the observable small scale density fluctuations,
in addition to the acoustic oscillation damping in existence of the baryon-dark matter coupling,
we apply the approximated treatment of dark matter free-streaming analogous to that of the conventional warm dark matter.
We also demonstrate the mass dependence of the baryon-dark matter cross section bounds
(for the dark matter mass down to $m_{\chi} \sim 5~{\rm keV}$),
in contrast to the dark matter mass independence of the cross section constraints
for the light dark matter below the MeV scale claimed in the previous literature.

\begin{description}
\item[PACS numbers]
95.35.+d ,98.80.Es
\end{description}
\end{abstract}
\pacs{Valid PACS appear here}

\maketitle

\section{\label{sec:level1}Introduction}

While there is a growing support for the existence of dark matter~(DM) from the astrophysical observables,
its precise nature such as its mass and how it interacts with the visible matter remains unknown. 
The current studies favor the cold dark matter~(CDM) paradigm \cite{Planck}
in which DM particles are ``cold''~(non-relativistic) and can interact with other particles only through the gravitational force.
While CDM can explain well the large-scale structure observations,
there are still unsettled challenges in the CDM paradigm on small scales
\cite{Kauffmann:1993gv,Flores:1994gz,Moore:1994yx,Moore:1999nt,Klypin:1999uc,
Read:2005zv,BoylanKolchin:2011de,Wetzel:2016wro,Kim:2017iwr} 
and there is hence still a room to seek the hint on the new physics beyond the CDM paradigm. 
We in this paper study the possibility for the DM non-gravitationally scattering with the baryons
and its consequences on the cosmological observables. 

Many theoretical works suggest the DM particle models which have non-gravitational interactions with baryons,
such as the DM interacting through its dipole moment (dipole DM) and the millicharged DM
\cite{Sigurdson:2004zp,dav,turnoff,essig2,eee,cora,kenjisn,celine,boesh,rocky,beam,bur,pos,dubo,Melchiorri:2007sq,mas,bank,ess,semi,light,barg2,heo3,ilidio,gary,kst,gk}
which have received the revived interests in view of the recent global 21~cm signal measurement by the EDGES
\cite{Bowman:2018yin,Tashiro:2014tsa,Munoz:2015bca,Barkana:2018lgd,Berlin:2018sjs,Kovetz:2018zan}.

Since the baryon-DM interactions can give the significant effects
on the fluid dynamics of primordial plasma and structure formations,
the observations of cosmic microwave background (CMB) anisotropies
and large-scale structure can tightly constrain the baryon-DM couplings
\cite{Loeb:2005pm,Bertschinger:2006nq,Dvorkin:2013cea}.
We in this paper study the constraints on such a DM model
by using the Planck CMB data ~\cite{Planck}
and the Lyman-$\alpha$ forest data from the Sloan Digital Sky Survey~(SDSS)~\cite{SDSS}.
Although there already exist recent works on this subject \cite{Xu:2018efh,Boddy:2018wzy,Slatyer:2018aqg},
our study extends the previous literature by covering the DM mass range below the MeV scale
and finds the unique features peculiar to the small mass
such as the dependence of the baryon-DM cross section bounds on the DM mass. 

The remainder of the paper is organized as follows:
Sec. II discusses the temperature and perturbation evolutions of the DM and baryon in existence of the baryon-DM coupling.
Sec. III describes our likelihood analysis for constraining the cosmological parameters including the baryon-DM coupling cross section,
followed by the presentation of the results in subsections along with the comparison to the previous works.
Finally, the summary of this work is given in Sec. IV.

\section{Cosmological evolution with baryon-dark matter coupling}

The baryon-DM coupling gives an impact on the thermal history and the
structure formation of the Universe. To evaluate such effects, we consider the velocity-dependent
baryon-DM scattering cross section parameterized as \cite{Dvorkin:2013cea,zurekyu,raby,plasma}
\beq
\label{eq:sigma}
\sigma(v) = \sigma_0 v^n,
\eeq
where $v$ denotes the baryon-DM relative velocity.
While $n$ depends on the type of the coupling,
we present our discussions for several representative values of $n$
without specifying the concrete particle theory models \cite{ArkaniHamed:2008qn, Buckley:2009in,Sigurdson:2004zp,Melchiorri:2007sq}
to keep our discussions as general as possible
so that our cosmological constraints can be applicable to a wide range of DM scenarios. We set $c=\hbar=k_{\rm B}=1$ in our discussions.

\subsection{Thermal evolution}
The evolution equations for the DM and baryon temperatures,
$T_\chi$ and $T_{\rm b}$ read, in existence with the baryon-DM coupling,
\begin{align}
\label{eq:temperature1}
\dot{T}_\chi = &-2\frac{H}{1+z} T_\chi + \frac{2m_\chi}{m_\chi+m_{\rm H} }
 R_\chi(T_{\rm b}-T_\chi), \\
\label{eq:temperature2}
\dot{T}_{\rm b} = &-2\frac{H}{1+z} T_{\rm b} + \frac{2\mu_{\rm b}}{m_\chi+m_{\rm H}}\frac{\rho_\chi}{\rho_{\rm b}}R_\chi(T_\chi-T_{\rm b}) \nonumber \\
&+ \frac{2\mu_{\rm b}}{m_{\rm e}}R_\gamma(T_\gamma-T_{\rm b}),
\end{align}
where the dot denotes a derivative with respect to the conformal time,
$\rho_\chi$ is the DM energy density, $\rho_{\rm b}$ is the baryon energy density,
and $m_\chi$, $m_{\rm H}$ and $m_{\rm e}$ are the DM, hydrogen and electron masses respectively.
$\mu_{\rm b} = m_{\rm H} (n_{\rm H}+4n_{\rm He})/(n_{\rm H}+n_{\rm He}+n_{\rm e})$ is the mean molecular weight
for baryons with the number density for hydrogen, helium and free electrons, $n_{\rm H},~n_{\rm He},~n_{\rm e}$,
and $R_\gamma = 4 n_{\rm e}\sigma_{\rm T} \rho_\gamma /{3\rho_{\rm b}}/(1+z)$ is the Thomson scattering rate
with the Thomson cross section $\sigma_{\rm T}$.
In Eq.~(\ref{eq:temperature1}), $R_\chi$ represents the baryon-DM scattering rate.
To calculate $R_\chi$, it is required to evaluate the relative velocity $v$
which is related not only to the thermal velocity dispersions of both baryons
and DM but also to the relative peculiar velocity between them.
Following Refs.~\cite{Dvorkin:2013cea,Xu:2018efh}, $R_\chi$ is given by 
\beq
\label{eq:rate_chi}
R_\chi = 
\frac{\rho_{\rm b}\sigma_0c_n}{m_\chi+m_{\rm H}}
\frac{{\mathcal F}_{\rm He}}{1+z}
\left( \frac{T_{\rm b}}{m_{\rm H}}+\frac{T_\chi}{m_\chi} +\frac{V^2_{\rm RMS}}{3} \right)^{\frac{n+1}{2}},
\eeq
where $V^2_{\rm RMS}$ is the rms peculiar velocity
and we adopt the evolution of $V^2_{\rm RMS}$ in the standard
cosmology \cite{Xu:2018efh},
\beq
\label{eq:v_rms}
V^2_{\rm RMS} = 
  \begin{cases}
    10^{-8}	& z > 10^3 \\
    10^{-8} \left( \frac{1+z}{10^3} \right)^2 & z \leq 10^3.
  \end{cases}
\eeq

In Eq.~(\ref{eq:rate_chi}), $c_n$ is the $n$-dependent constant
and ${\mathcal F}_{\rm He}$ represents a ratio of the cross section for helium to hydrogen
to take account of the difference in DM-hydrogen and DM-helium scatterings \cite{Dvorkin:2013cea}.
In our calculation, following previous works \cite{Dvorkin:2013cea,Xu:2018efh},
both $c_n$ and ${\mathcal F}_{\rm He}$ are set to
$c_n \approx  \{0.27,0.53,1,2.1,13\}$ with $n =\{-4,-2,-1,0,2\}$
and ${\mathcal F}_{\rm He}=0.76$ for simplicity.

The tight coupling can hold roughly until the redshift $z_{\rm dec}$
satisfying $R_\chi {m_\chi}/({m_\chi+m_{\rm b}}) = H_{\rm dec}/(1+z_{\rm dec})$
where the subscript $\rm dec$ represents the value at $z_{\rm dec}$.
$(1+z)R_\chi/H$ can increase for a bigger $z$ if $n>-3$,
and the DM temperature can tightly couple with the baryon temperature
at a high redshift for $n>-3$ \cite{Dvorkin:2013cea}.
We hence adopt the sudden decoupling approximation in treating the thermal evolution, for $n>-3$, until $z=10^4$
for the case with $z_{\rm dec}>10^4$ in order to reduce the computational cost\footnote{This
"sudden decoupling approximation" induces a negligible difference
in our calculations for the cross section bounds compared with the exact ones \cite{Xu:2018efh}.},
\beq
\label{eq:T_ini}
T_\chi =
  \begin{cases}
    T_{\rm b},	& R_\chi\frac{m_\chi}{m_\chi+m_{\rm b}} >  \frac{H}{1+z}\\
    T_{\rm dec}\left( \frac{1+z}{1+z_{\rm dec}} \right)^2, & R_\chi\frac{m_\chi}{m_\chi+m_{\rm b}} < \frac{H}{1+z}.
  \end{cases}
\eeq
For $z<10^4$, the temperature evolution is obtained from
Eqs.~(\ref{eq:temperature1}) and~(\ref{eq:temperature2}) numerically with the initial condition Eq.~(\ref{eq:T_ini}).

On the other hand, when $z_{\rm dec}<10^4$, we take the tight thermal coupling $T_{\rm b}=T_{\chi}$
before $z=10^4$ and we thereafter numerically solve Eqs.~(\ref{eq:temperature1}) and~(\ref{eq:temperature2}).

For $n=-4$, since the coupling between baryon and DM components is weaker at a higher redshift,
the DM temperature evolves adiabatically,~$T_\chi \propto (1+z)^2$,
after becoming non-relativistic and is much lower than the CMB and baryon temperatures.
We therefore solve Eqs.~(\ref{eq:temperature1}) and (\ref{eq:temperature2}) from $z=10^4$ with the initial condition $T_\chi=0$.

\begin{figure}[ht]
\includegraphics[width=9cm,]{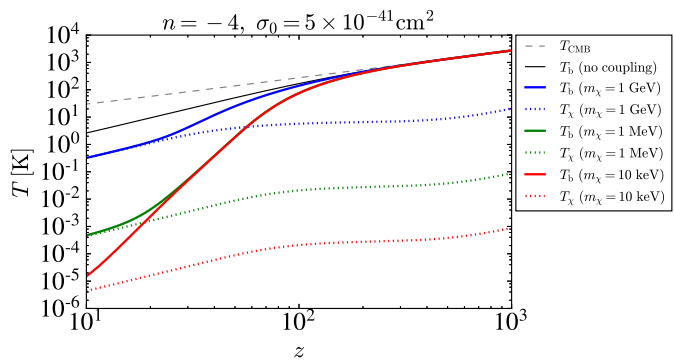}
\caption{\label{fig:temperature} The time evolutions of photon, baryon and DM temperatures.}
\end{figure}

For an illustration purpose, we show the evolutions of photon, baryon and DM temperatures
with $n=-4$ for different DM masses in Fig.~\ref{fig:temperature}.
The low DM temperature can be heated up by the higher baryon temperature due to the baryon-DM couplings.
We can see in this figure that, after the baryon temperature decoupled from the CMB temperature,
the baryon temperature is dragged toward the DM temperature and, resultantly,
the DM and baryon temperatures can couple at low redshifts.

\subsection{Density fluctuation evolution}

We now consider the evolution of the density fluctuations in existence of the baryon-DM coupling.
We work in the Newtonian gauge in the following.
For a given Fourier mode $k$, the density fluctuations, $\delta_{\rm b}$ and $\delta_\chi$,
and the divergences of the fluid velocities\footnote{Note that $\theta = ikv$, here $v$ is the fluid velocity.},
$\theta_{\rm b}$ and $\theta_\chi$,
for baryon and DM components evolve as
\begin{align}
\label{eq:perturbation}
\dot{\delta}_{\rm b} = &-\theta_{\rm b} + 3\dot{\phi}, \quad \dot{\delta}_\chi = -\theta_\chi + 3\dot{\phi}, \\
\dot{\theta}_\chi = &-\frac{H}{1+z}\theta_\chi + c_\chi^2k^2\delta_\chi + R_\chi(\theta_{\rm b} - \theta_\chi) + k^2\psi, \\
\dot{\theta}_{\rm b} = &-\frac{H}{1+z}\theta_{\rm b} + c_{\rm b}^2k^2\delta_{\rm b}
+ \frac{\rho_\chi}{\rho_{\rm b}}R_\chi(\theta_\chi - \theta_{\rm b}) \nonumber\\
&+ R_\gamma(\theta_\gamma - \theta_{\rm b}) + k^2\psi,
\label{eq:perturbation-4}
\end{align}
where $\psi$ and $\phi$ are the Newtonian gravitational potential and the spatial metric perturbation, respectively.
$c_{\rm b}$ and $c_\chi$ represent the sound speeds of each fluid
\begin{align}
\label{eq:sound_speed}
c^2_{\rm b} &= \frac{T_{\rm b}}{\mu_{\rm b}} \left( 1 + \frac{1+z}{3}\frac{d\ln T_{\rm b}}{dz} \right), \\
c^2_\chi &= \frac{T_\chi}{m_\chi} \left( 1 + \frac{1+z}{3}\frac{d\ln T_\chi}{dz} \right).
\end{align}

For the initial condition of the fluctuations, we adopt the standard
adiabatic initial condition in which the initial density fluctuations
and the velocity divergence match the baryon ones.

We focus on the DM mass below the MeV scale.
In such a mass range, we have to take into account a relativistic free-streaming effect
which erases small scale density fluctuations in the early universe,
as in the analysis for the warm dark matter~(WDM) model.

The public Boltzmann code,~{\tt CLASS}, provides us the treatment to solve
the Boltzmann equations for the WDM model, and we modified the {\tt CLASS}
in order to calculate
Eqs.~\eqref{eq:perturbation}--\eqref{eq:perturbation-4}.
This method is however computationally demanding in performing the Markov chain Monte Carlo~(MCMC) analysis
to obtain the observational constraint on the baryon-DM coupling parameters.
To facilitate our analysis (without losing much precision as demonstrated in the following section),
we adopt the following WDM approximation to take into account the free-streaming effect due to the small DM mass.

\begin{figure}[ht]
\includegraphics[width=7.5cm,]{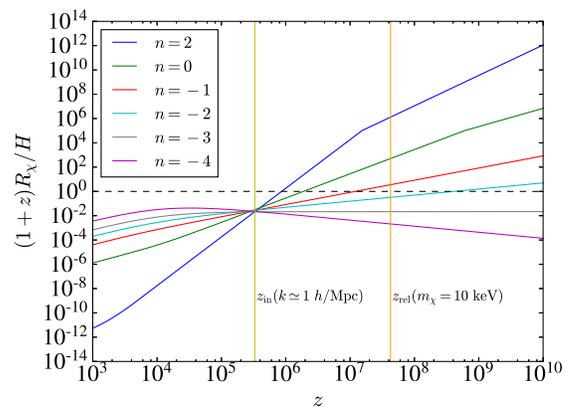}
\caption{\label{fig:Rchi_H}
The time evolution of the ratio between two time scales, $R_\chi$ and $H/(1+z)$.
$z_{\rm rel}$ (orange vertical line) is the redshift before which the DM with $m_\chi=10\ {\rm keV}$ is relativistic.
$z_{\rm in}$ (yellow vertical line) is the redshift
when the scale $k=1\ h/$Mpc (relevant to the Ly-$\alpha$ data) enters the horizon.}
\end{figure}

\subsection{The matter power spectrum suppression}

The matter power spectrum suppression due to the acoustic damping can affect the Ly-$\alpha$ observations
if the DM couples to the baryon around $z\sim 10^6$.
The mode relevant for the Ly-$\alpha$ observations ($k\sim 1\ h$/Mpc) enters the horizon around $z\sim 10^6$
(corresponding to the temperature of order keV),
and the DM hence can be treated as non-relativistic around $z\sim 10^6$
of our interest if the mass is above keV. 

In addition to such acoustic oscillation damping effects,
the DM free-streaming effects can also suppress the matter power spectrum.
The DM can free-stream once it decouples from the baryons,
and the matter power spectrum due to such free-streaming effects can potentially appear
at the small scales observable by the Ly-$\alpha$ measurements if the DM is relativistic at the decoupling epoch.
If the DM is non-relativistic at the decoupling, on the other hand,
the free-streaming length would be too small to be observable by the Ly-$\alpha$. 
In our computation, we regard the free-streaming effect is negligible
when the decoupling temperature become $T_{\rm dec} < 0.1\ m_\chi$.

To illustrate the DM decoupling epoch and the epoch when DM becomes non-relativistic,
we plot the ratio of two time scales, $R_\chi$ and $H/(1+z)$,
in Fig.~\ref{fig:Rchi_H} \footnote{The sudden decoupling approximation is reflected on the slight kink in the scattering rate evolution.}.
The orange vertical line represents the redshift, $z_{\rm rel}$,
before which the DM with mass $m_\chi=10\ {\rm keV}$ is relativistic.
The yellow vertical line is the redshift $z_{\rm in}$ around
which the largest observable wavenumber $k= 1~\rm Mpc^{-1}$ enters the horizon.

If the ratio of two scales $(1+z)R_\chi/H$ is much larger than unity, the baryon-DM coupling is effective.\\

Let us now outline our treatment of the free streaming effects.

First, for $n \geq -3$, we solve Eqs.~\eqref{eq:perturbation}--\eqref{eq:perturbation-4}
numerically without the free-streaming effect
and obtain the power spectrum $P_*(k)$ in Eq.~\eqref{eq:wdm_approx}.
Then, similar to the manner in the warm DM analysis~\cite{Bode:2000gq},
we postprocess the free-streaming effect on the power spectrum
by using the transfer function~${\cal T}(k)$.
\beq
{\cal T}(k) = \left[ 1+(\alpha k)^{2\nu} \right]^{-5/\nu},
\eeq
with the fitting parameters, $\alpha$ and $\nu$.
We adopt the fitting parameters suggested in Ref.~\cite{Viel:2005qj};
\beq
\alpha = 0.24\left( \frac{m_\chi/T_{\chi}}{1{\rm keV}/T_{\nu}} \right)^{-0.83}
\left( \frac{\Omega_\chi h^2}{0.12} \right)^{-0.16} {\rm Mpc},
\label{eq:alpha}
\eeq
and $\nu = 1.12$.
$\Omega_\chi h^2$ is the current value of the DM density parameter
multiplied by the square of the Hubble constant $H_0$ (here, $H_0 = 100h\ {\rm km}\, {\rm s}^{-1}{\rm Mpc}^{-1}$).
The temperature
ratio between the DM and the neutrino components is written as \cite{Viel:2005qj}
\beq
\frac{T_{\chi}}{T_{\nu}} = \left( \frac{10.75}{g_*(T_{\rm dec})} \right)^{1/3},
\eeq
where $g_*(T_{\rm dec})$ denotes degrees of freedom at temperature $T_{\rm dec}$ when the DM decoupling occurs.
We compute $g_*(T_{\rm dec})$ by using Table A1 in Ref.~\cite{Husdal:2016haj}.

In short, our final matter power spectrum with considering the free-streaming effect of the light DM mass is given by
(compared with the standard CDM matter power spectrum $P_*$ without the free streaming effects)
\beq
P(k) ={\cal  T}^2(k) P_*(k).
\label{eq:wdm_approx}
\eeq

\begin{figure}[ht]
\includegraphics[width=8cm,]{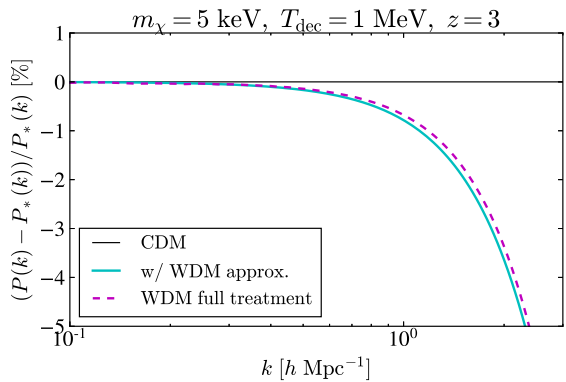}
\caption{\label{fig:wdm_pk}
The matter power spectrum difference from the standard CDM case at the redshift $z=3$ are shown.
The cyan solid curve shows the one evaluated from Eq.~\eqref{eq:wdm_approx},
while the magenta dashed curve shows the full treatment of the WDM case with $m_\chi=5\ {\rm keV}$,
$T_{\rm dec}=1\ {\rm MeV}$.}
\end{figure}

Fig. \ref{fig:wdm_pk}  shows the effects of free streaming and compare the power spectra
between the one evaluated from Eq.~\eqref{eq:wdm_approx}
and the one computed with the full numerical treatment by the {\tt CLASS}.
We can see the discrepancy is less than 1 \% for the parameter range of our interest,
and our using the fitted transfer function instead of using the computationally demanding full numerical solution
suffices for our purpose of demonstrating the possible bounds on the baryon-DM cross sections.

We note that such a free-streaming effect approximation discussed above however does not work for $n<-3$.  
As shown in Fig.~\ref{fig:Rchi_H}, the baryon-DM coupling becomes weaker as the redshift increases.
In this case, the DM particles are not thermally coupled with the baryon in the early epoch.
One expects that the baryon-DM interaction cross section can be more tightly constrained by the CMB
than by the Ly-$\alpha$ for $n = -4$ because the baryon-DM interactions become stronger (enhanced by the quartic power of the velocity) at the epoch $z\sim 10^3$
relevant for the CMB rather than $z\sim 10^6$ relevant for Ly-$\alpha$.
Hence, even though the treatment of the non-thermally produced DM is model-dependent,
we simply consider the parameter ranges where the dispersion of our DM is too small to be relevant for the Lyman-$\alpha$
observations\footnote{Even though we do not specify a concrete model for the purpose of general discussions,
the neglection of the free streaming effects for $n=-4$ can be applicable, for instance,
when the cold (thermally decoupled) parent particle with the mass $M$ decays into the lighter DM at $T=T_{\rm dec}$ for $M \ll T_{\rm dec}$.
The dispersion scales as the momentum over the mass \cite{bond1980,colo1995,Bode:2000gq,Viel:2005qj},
and such a scenario can lead to a much smaller DM velocity dispersion scale
than the conventional warm DM scenario at least by a factor $(M/T_{\rm dec}) (m_{\rm WDM}/m_{\chi})$.
See, for instance, Refs. \cite{boya2008,Bae:2017dpt} for the calculation of the phase space density
and the resultant velocity dispersion for the non-thermally produced DM.}.

\begin{figure}[ht]
\includegraphics[width=7cm,]{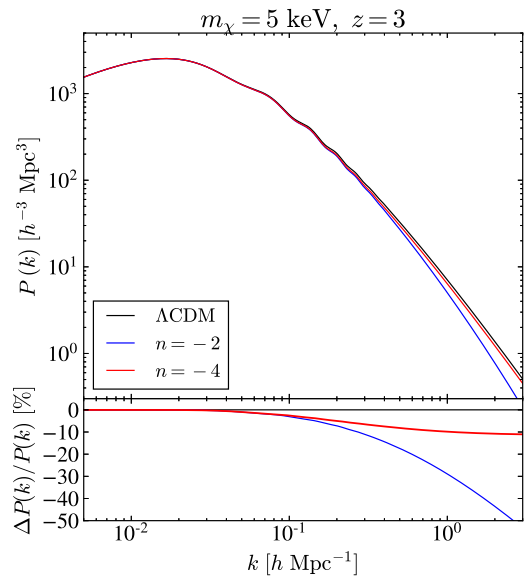}
\caption{\label{fig:Pk} Matter power spectra for the standard CDM model (black)
and that including the baryon-DM couplings for $n=-2$ (blue) and $-4$ (red).
The baryon-DM cross section constant $\sigma_0$ is set to the 95\% C.L. upper bound values
tabulated in Table~\ref{tab:table1}.}
\end{figure}

\begin{figure}[ht]
\includegraphics[width=7cm,]{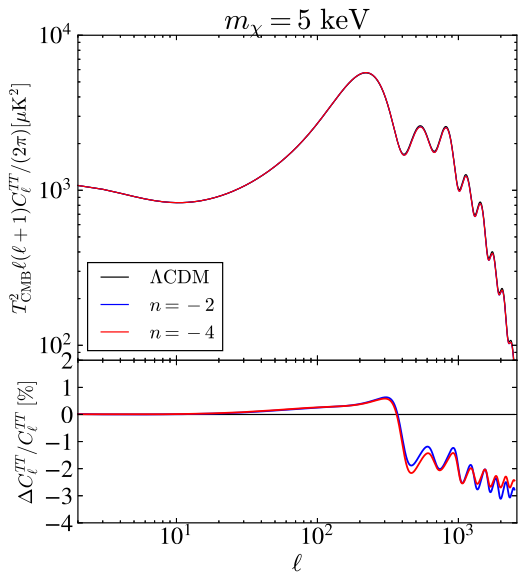}
\caption{\label{fig:Cl} The CMB temperature angular power spectra for the standard CDM model (black)
and those including the baryon-DM couplings for $n=-2$ (blue) and $-4$ (red).
The baryon-DM cross section constant $\sigma_0$ is set to the 95\% C.L. upper bound values
tabulated in Table~\ref{tab:table1}.
The other cosmological parameters are fixed to the standard values.}
\end{figure}

The typical matter power spectra and angular power spectra of the CMB temperature fluctuations
are shown in Figs. \ref{fig:Pk} and \ref{fig:Cl} where the baryon-DM cross section
is set to the upper bound values allowed by the 95\% C.L. limits.
We now discuss how these bounds on the cross sections are obtained.

\section{constraint}

Our final aim is to provide the constraint on the baryon-DM coupling
for the DM mass below 1~MeV scale.
After outlining our analysis method in Sec.~\ref{sec:mcmc},
we show the constraint from the CMB alone in Sec.~\ref{sec:cmb} which should be compared with the limits
by adding Ly-$\alpha$ data in Sec.~\ref{sec:lya}.

\subsection{\label{sec:mcmc}MCMC analysis}

To compute the temperature and polarization fluctuations in the CMB and the matter power spectrum, 
we numerically solve the Boltzmann equations including the baryon-DM coupling
by modifying the publicly available numerical code,
{\tt CLASS}~\cite{CLASS2} as described in the previous section.

Our analysis uses the CMB angular power spectrum data (TT + lowP + lensing) from the Planck ~\cite{Planck}
and the Lyman-$\alpha$ flux power spectrum from SDSS~\cite{SDSS} which
gives the matter power spectrum at redshift $z=3$ around $k\simeq1\ h\,{\rm Mpc^{-1}}$.
In this paper, we adopt the MCMC method
with Monte Python~\cite{Audren} developed in the {\tt CLASS} code.

For the MCMC analysis, we have seven free parameters.
The free parameter for the baryon-DM coupling
is only $\sigma_0$, while we fix the DM mass~$m_\chi$ and the spectral index for
the velocity dependence,~$n$.
For the other six free parameters,
we set the priors for the standard cosmological parameters as
\begin{align}
\label{eq:prior}
100\Theta \in (0.5,10),\ \ \Omega_{\rm b} h^2 \in (0.005,0.04), \nonumber\\
\Omega_\chi h^2 \in (0.01,0.5),\ \ \tau_{\rm reio} \in (0.005,0.5), \\
{\rm ln}(10^{10}A_{\rm s}) \in (0.5,10),\ \ n_{\rm s} \in (0.5,1.5). \nonumber
\end{align}
where $\Theta$ is the angular size of the sound horizon at recombination,
$\Omega_{\rm b} h^2$ and $\Omega_\chi h^2$ are the current values of the baryonic and DM density parameters
multiplied by the square of the Hubble constant $H_0$ (here, $H_0 = 100h\ {\rm km}\, {\rm s}^{-1}{\rm Mpc}^{-1}$),
$\tau_{\rm reio}$ is the reionization optical depth and $A_{\rm s}$ and $n_{\rm s}$
are the amplitude and spectral index of the primordial power spectrum.

\subsection{\label{sec:cmb}The bounds from the CMB data alone}

Let us show the results of the parameter constraints.
First we present the limits from the MCMC analyses by using the CMB data alone.
Such analysis using only the CMB data have been performed in the previous works
\cite{Dvorkin:2013cea,Xu:2018efh,Boddy:2018wzy},
but our study extends the previous works by covering the DM mass lighter ($\lesssim 10$keV)
than that discussed in the the previous literature
and found the unique features peculiar to such small DM masses.

\begin{figure}[hb]
\begin{tabular}{cc}
\includegraphics[width=4.05cm,]{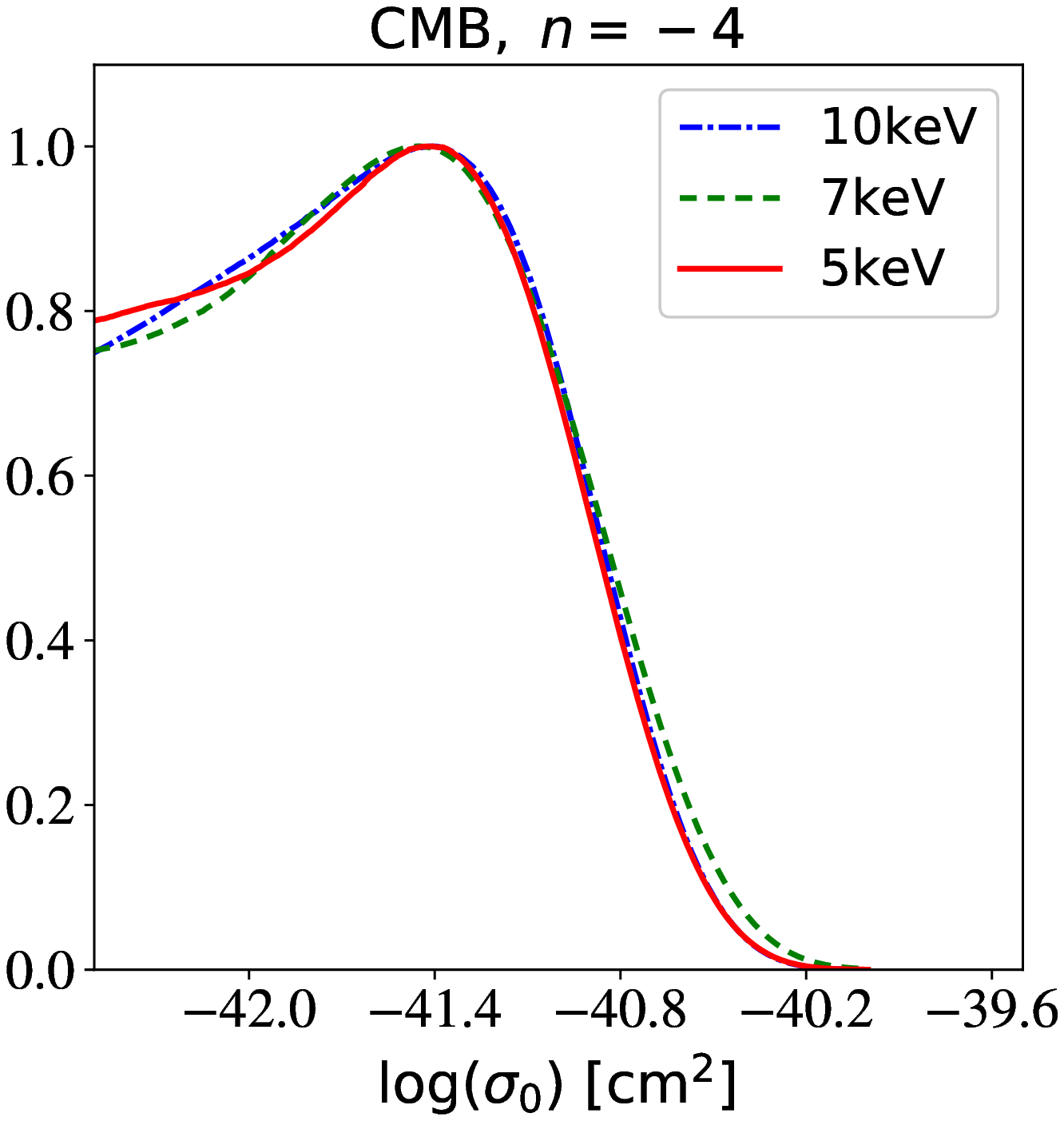}&
\includegraphics[width=4.2cm,]{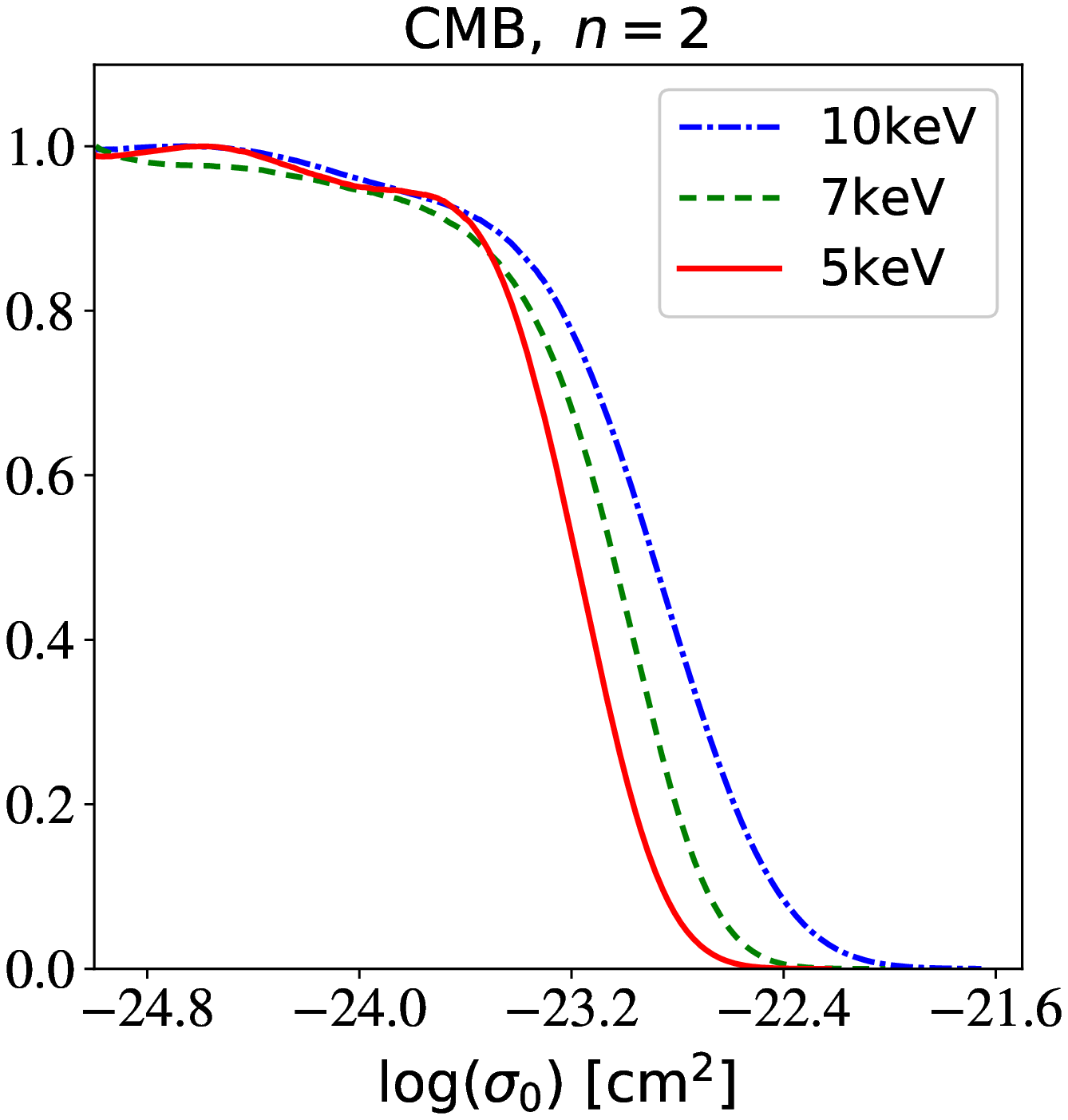}\\
\end{tabular}
\caption{\label{fig:ba_sigma}
The posterior distribution of $\sigma_0$.
Here we use the CMB data alone for the MCMC analyses.
The other parameters are marginalized over.}
\end{figure}

Table~\ref{tab:table1} lists the results of the constraint
on the cross section of the baryon-DM coupling $\sigma_0$
from the MCMC analyses by using the CMB data alone.
We show the marginalized posterior distribution of $\sigma_0$ in Fig.~\ref{fig:ba_sigma}.
In the previous works,
Ref.~\cite{Xu:2018efh} has investigated
the constraint with the mass range $m_\chi \geq10\ {\rm MeV}$
for $-4\leq n \leq 2$ cases by using the CMB+Ly$\alpha$ data
and Ref.~\cite{Boddy:2018wzy} has studied the baryon-DM coupling
with the mass range~$m_\chi \geq10\ {\rm keV}$ for $n=-4,-2$ cases by using the CMB data alone.
Their works show that, when $m_\chi\ll 1\ {\rm GeV}$,
the constraint on $\sigma_0$ becomes independent of the DM mass.
As shown in Table~\ref{tab:table1} and Fig.~\ref{fig:ba_sigma},
our constraint is consistent with them and
we confirmed that there is no mass-dependence in the constraints
even for the mass range,~$\leq10\ {\rm keV}$,
except for the $n=+2$ case which gives a unique feature for a small DM mass.

This interesting mass dependence for $n=2$ which has not been found in the previous works 
shows up because of the strong $n$-dependence of the coupling rate.
The smaller the DM mass is,
the larger the $T_\chi/m_\chi$ term becomes in Eq.~(\ref{eq:rate_chi}).
A consequent enhancement of the baryon-DM interaction can result in
the tighter bounds on $\sigma_0$ as the DM mass decreases.
This characteristic feature becomes more apparent for a bigger $n$
because of the bigger $(n+1)/2$ power dependence of $R_{\chi}$
and a bigger value of $T_{\chi}$ in Eq.~(\ref{eq:rate_chi}),
and it is most prominent for $n=2$ compared with the other smaller values of $n$.
For instance, for $n=-4$, the DM decoupled sufficiently early such that
$T_{\chi}$ is too small to see the appreciable dependence of $\sigma_0$ on $m_{\chi}$.

\begin{table}[ht]
\caption{\label{tab:table1}
$95\%$ confidence limits for the upper bounds
on the baryon-DM cross section constant $\sigma_0$ in units of $\rm cm^2$,
where only the CMB data was used in the analysis.}
\begin{ruledtabular}
\begin{tabular}{cccc}
&\multicolumn{3}{c}{CMB ($95\%$\ C.L.)}\\
\colrule
\textrm{$n$}&
\textrm{$m_\chi =5$ keV}&
\textrm{$m_\chi =7$ keV}&
\textrm{$m_\chi =10$ keV}\\
\colrule
$-4$ & $1.6 \times 10^{-41}$ & $1.6 \times 10^{-41}$ & $1.6 \times 10^{-41}$\\
$-2$ & $2.0 \times 10^{-33}$ & $2.0 \times 10^{-33}$ & $2.0 \times 10^{-33}$\\
$-1$ & $1.1 \times 10^{-29}$ & $1.1 \times 10^{-29}$ & $1.1 \times 10^{-29}$\\
$0$ & $3.4 \times 10^{-26}$ & $3.4 \times 10^{-26}$ & $3.7 \times 10^{-26}$\\
$+2$ & $7.1 \times 10^{-24}$ & $1.0 \times 10^{-23}$ & $1.6 \times 10^{-23}$
\end{tabular}
\end{ruledtabular}
\end{table}

\begin{table}[ht]
\caption{\label{tab:table2}
$95\%$ confidence limits for the upper bounds on the $\sigma_0$ in units of $\rm cm^2$,
where the CMB and Ly$-\alpha$ data were used.}
\begin{ruledtabular}
\begin{tabular}{cccc}
&\multicolumn{3}{c}{CMB + Lyman-$\alpha$ ($95\%$\ C.L.)}\\
\colrule
\textrm{$n$}&
\textrm{$m_\chi =5$ keV}&
\textrm{$m_\chi =7$ keV}&
\textrm{$m_\chi =10$ keV}\\
\colrule
$-4$ & $1.8 \times 10^{-41}$ & $1.9 \times 10^{-41}$ & $1.9 \times 10^{-41}$\\
$-2$ & $1.8 \times 10^{-34}$ & $1.9 \times 10^{-34}$ & $2.2 \times 10^{-34}$\\
$-1$ & $2.9 \times 10^{-31}$ & $3.6 \times 10^{-31}$ & $3.6 \times 10^{-31}$\\
$0$ & $3.2 \times 10^{-29}$ & $4.8 \times 10^{-29}$ & $6.7 \times 10^{-29}$\\
$+2$ & $4.0 \times 10^{-31}$ & $1.2 \times 10^{-29}$ & $3.9 \times 10^{-28}$
\end{tabular}
\end{ruledtabular}
\end{table}

\begin{figure}[ht]
\begin{tabular}{cc}
\includegraphics[width=4cm,]{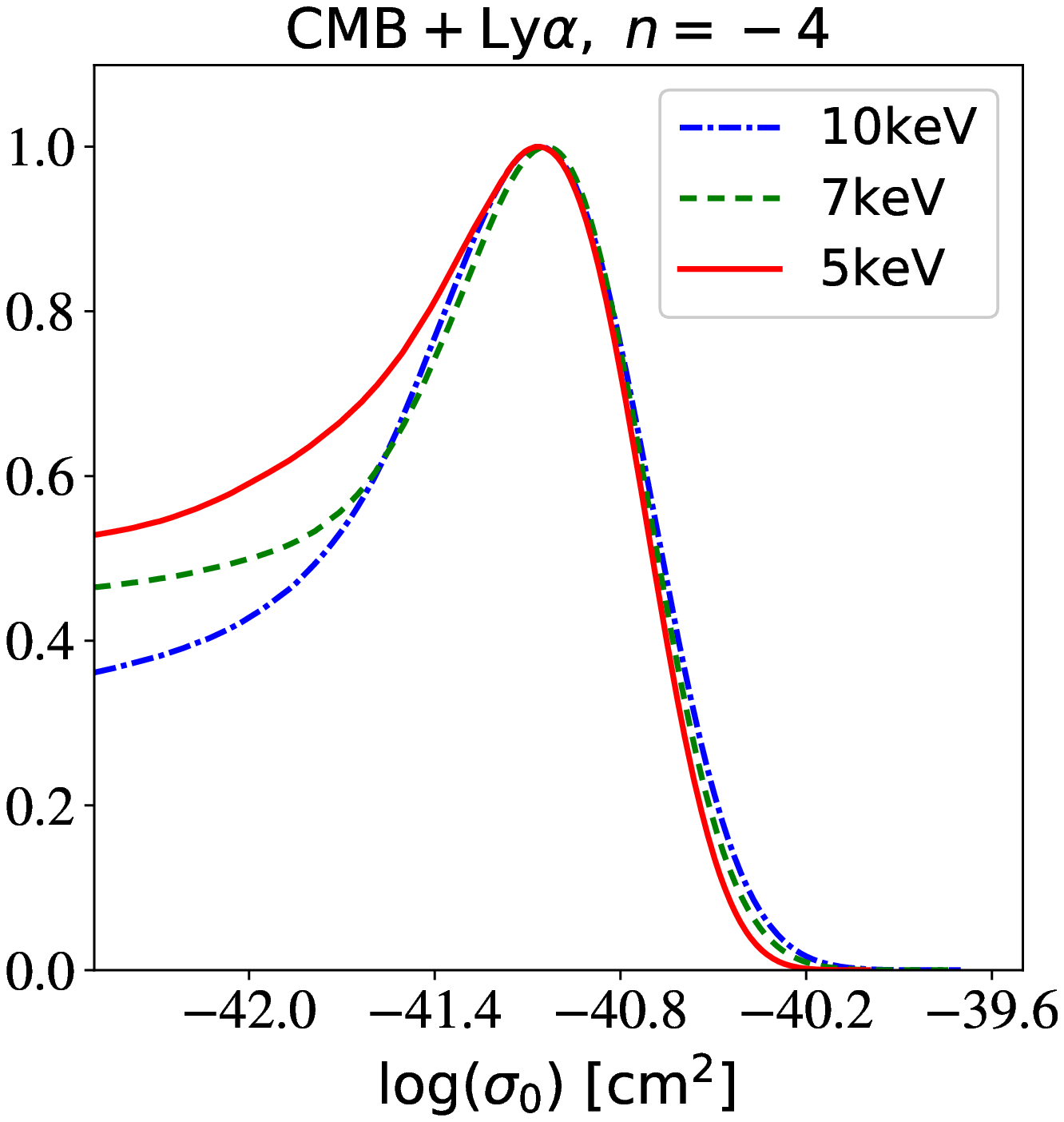}\\
\includegraphics[width=4cm,]{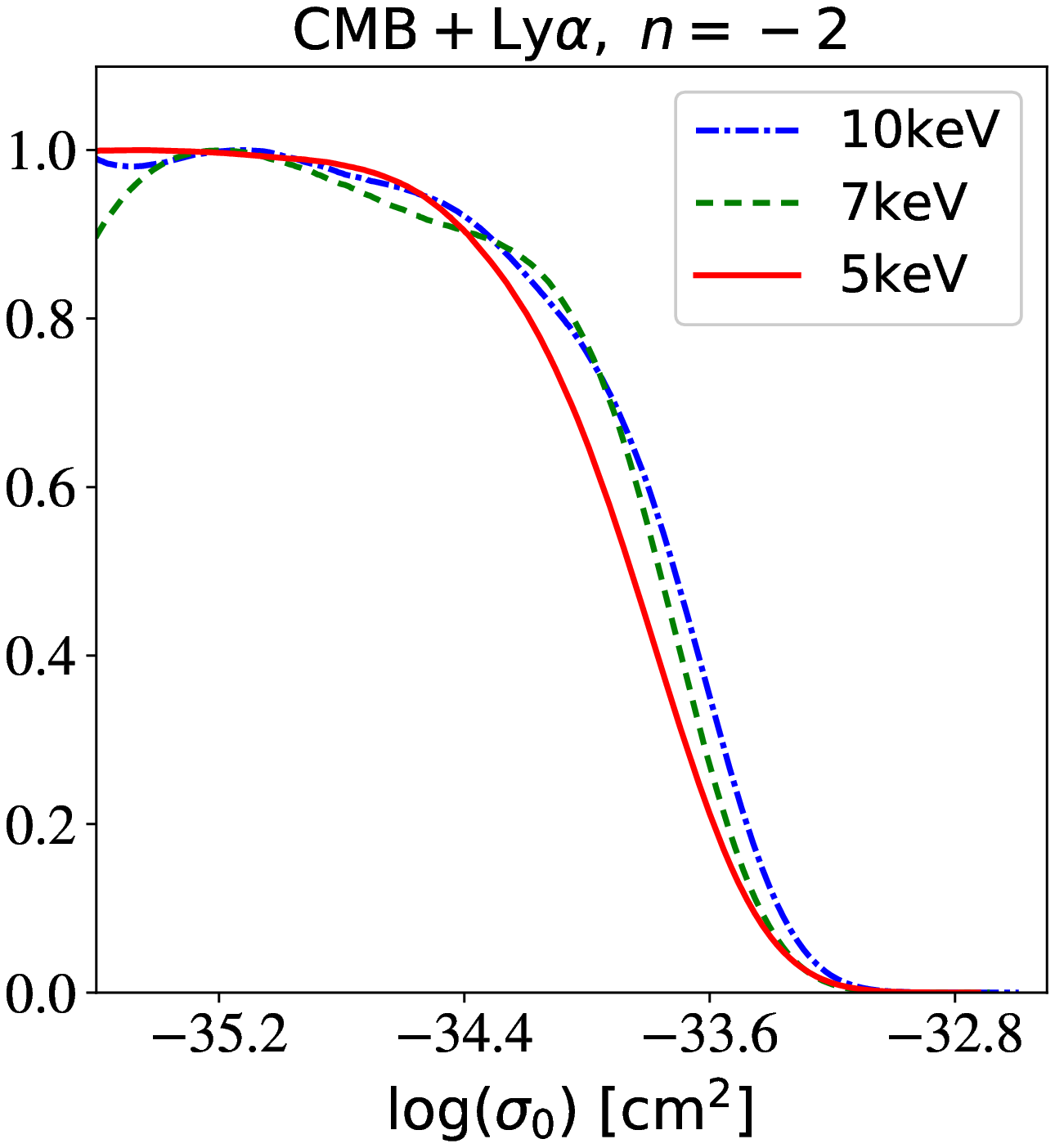}&
\includegraphics[width=4.2cm,]{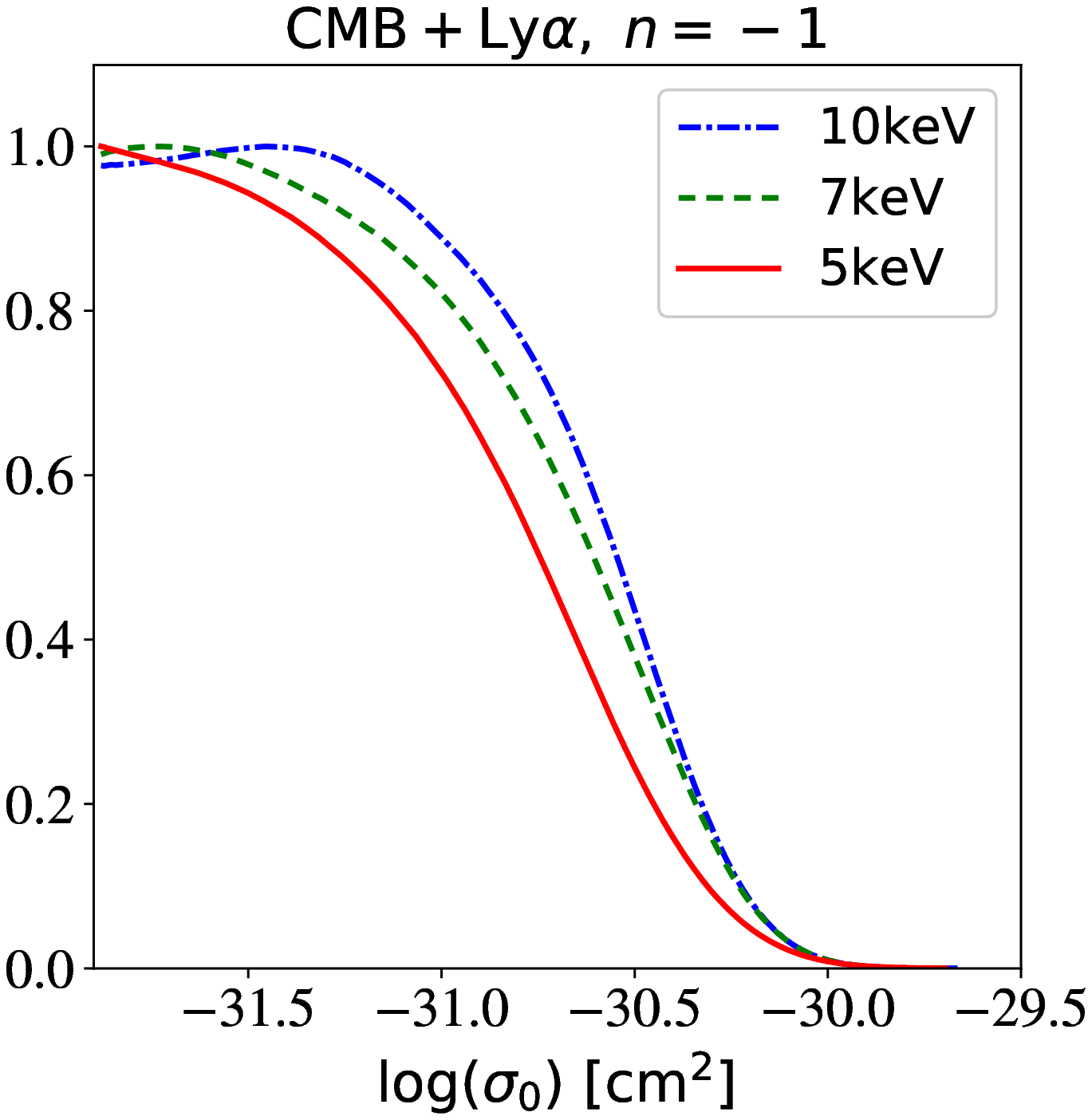}\\
\includegraphics[width=4.3cm,]{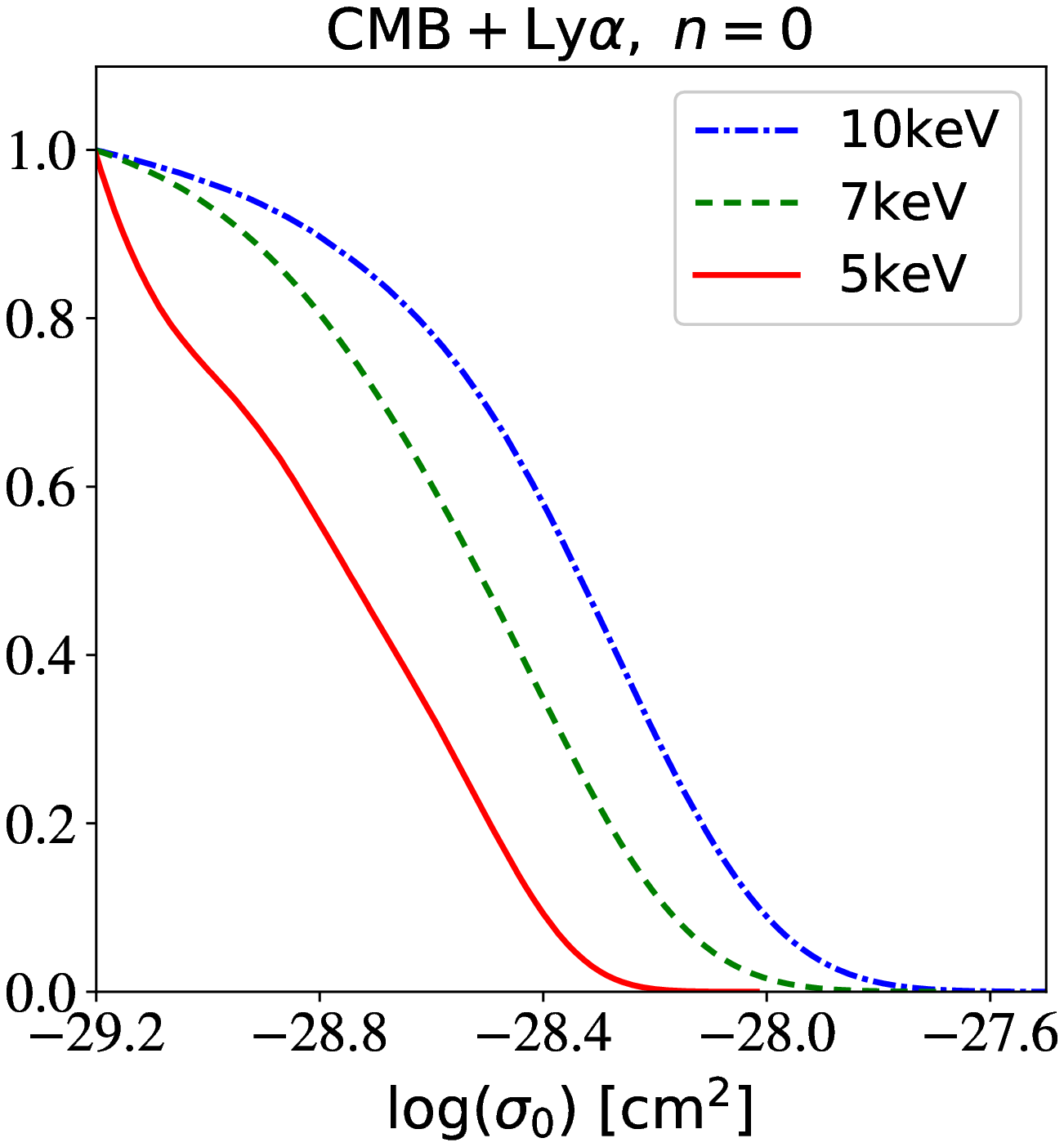}&
\includegraphics[width=4.2cm,]{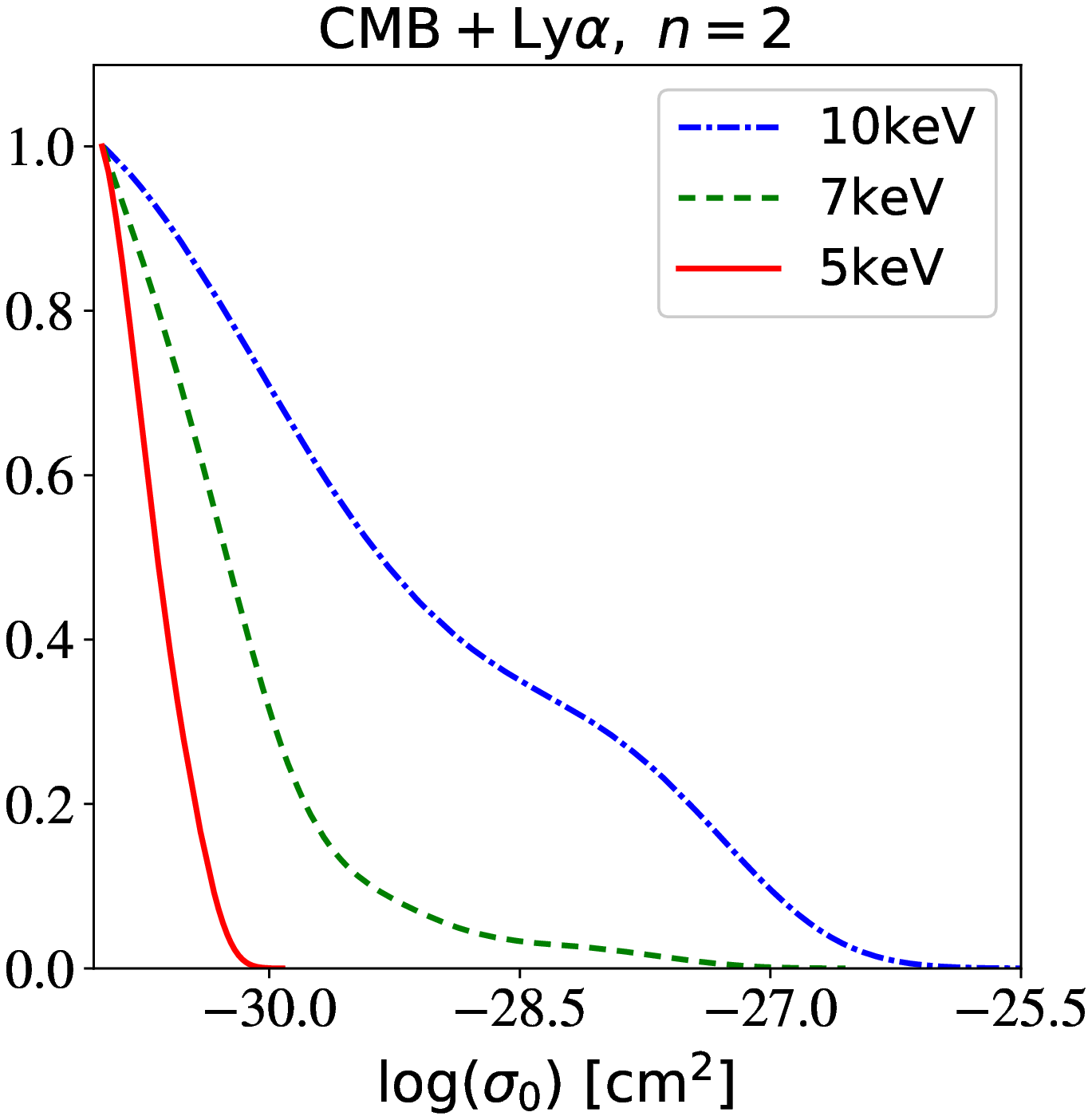}\\
\end{tabular}
\caption{\label{fig:ly_sigma}
The posterior distribution of $\sigma_0$ using the CMB and Ly$\alpha$ data in the MCMC analyses.
The other parameters are marginalized over.}
\end{figure}

\begin{figure}[ht]
\includegraphics[width=8cm,]{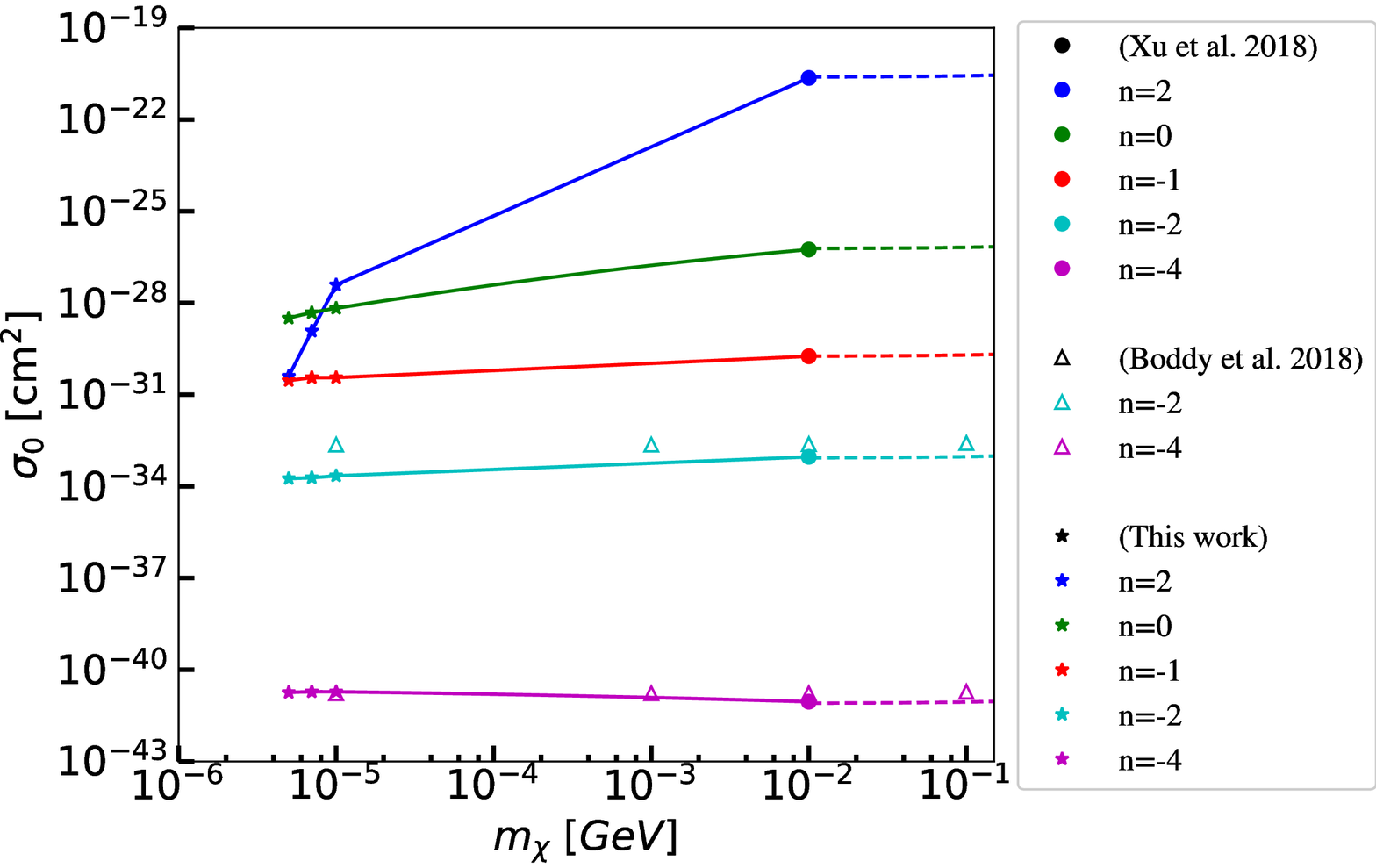}
\caption{\label{fig:sigma_m}
The constrained region at 95 \% C.L. on the $\sigma_0 - m_\chi$ plane.
The stars show the results of our work,
while the filled circles and the dashed lines show the results of Ref.~\cite{Xu:2018efh} using the CMB + Ly$\alpha$ data
and the triangles show the results of Ref.~\cite{Boddy:2018wzy} using the CMB data alone.
Our results and previous ones are extrapolated by the solid lines for the illustration purpose.
Our analysis gives the tighter constraints with the mass dependence than the previous works
except for $n=-4$ (which is dominated by the CMB constraints as discussed in Sec. III).}
\end{figure}

\subsection{\label{sec:lya}The bounds from CMB + Lyman-$\alpha$}

We next show the constraint from
the CMB + Lyman-$\alpha$ data, which is summarized in Table~\ref{tab:table2}.
The marginalized posterior distribution of $\sigma_0$ is shown in Fig.~\ref{fig:ly_sigma}.

As the DM mass becomes sufficiently small ($m_{\chi}\lesssim 1$ MeV),
the suppression of the power spectrum due to the free-streaming can become important
on the small scales depending on $n$ values
as discussed in the previous section for some parameter range of our interest.
Therefore, the additional suppression
due to the baryon-DM coupling can be tightly constrained, except for the $n=-4$ case.
We find that the combination of CMB + Lyman-$\alpha$ can provide the stronger
constraint than that from the CMB alone with $n>-3$.
This is reasonable because $R_{\chi}/aH$ becomes bigger for a bigger redshift for $n>-3$,
which gives the stronger suppression due to the baryon-DM coupling
in the redshift $z\sim 10^6$ when the modes relevant for the Lyman-$\alpha$ ($k\simeq1\ h\,{\rm Mpc^{-1}}$) enter the horizon.
For $n=-4$, on the other hand, $R_{\chi}/aH$ increases for a smaller redshift
and the baryon-DM momentum transfer rate
becomes sufficiently large at the CMB epoch ($z\sim 10^3$),
and we expect to get the tight constraint from the CMB rather than from the Lyman-$\alpha$.
The comparison of Fig. \ref{fig:ba_sigma} with Fig. \ref{fig:ly_sigma} for $n=-4$
indeed shows no appreciable difference in the best-fit values for $\sigma_0$
and the inclusion of the Lyman-$\alpha$ data does not give us a significant improvement in constraining $\sigma_0$.
For the $n=-4$ case, the marginalized posterior distribution in Fig. \ref{fig:ly_sigma} shows that
the MCMC analysis prefers a somewhat larger value of the cross section $\sigma_0$
than that preferred by the analysis including only the CMB data.
We point out that this is due to the degeneracy between $n_{\rm s}$ and $\sigma_0$.
The analysis tries to fit the deficit of the CMB power spectrum on large scales by a larger $n_{\rm s}$
which can suppress the power below the pivot scale $k\leq 0.05\ h$/Mpc.
Such a larger $n_{\rm s}$ gives the excess on small scales ($k\geq 0.05\ h$/Mpc),
and, to cancel this excess, a large $\sigma_0$
(which can suppress the small scale power without affecting the large scale) is preferred.

Fig.~\ref{fig:sigma_m} illustrates our discussions in this section
and summarizes the results of the constraints on the $\sigma_0 - m_\chi$ plane along with results from the previous works.

\section{Summary}
We have investigated the cosmological constraints
on the cross section of a baryon-DM coupling
which is parameterized as the power of the relative velocity between DM and baryons.
This parameterization allows us to generally treat many DM models
which possess the couplings with baryons.
We have performed the MCMC analysis
by using the angular power spectrum of the CMB from the Planck and the small-scale matter power spectrum
obtained from the Lyman-$\alpha$ forest data by the SDSS.
Our obtained result includes the new constraint
for a smaller DM mass below the MeV scale and a wider range of the power
law index, $-4\leq n\leq 2$, compared with the previous works.

Even though we explored the DM mass range only down to $m_{\chi} \sim 5$ keV
because of the uncertainty due to the treatment of the collision term for the relativistic species,
we already could start seeing interesting characteristic features
such as the potentially strong mass dependence of the momentum transfer rate
and hence that of the cross section bounds so that the further exploration
of the even lighter DM mass range than studied here would be warranted.
Our result should offer a complementary study to other studies
including other astrophysical constrains and DM direct detection bounds \cite{Sigurdson:2004zp,Xu:2018efh}.
For instance, the Lyman-$\alpha$ forest observation by the Dark Energy Spectroscopic Instrument (DESI)~\cite{DESI}
could be of great interest to get the stronger constraint on the baryon-DM coupling, which is left for our future work.

\begin{acknowledgments}
We thank J. Silk for the useful discussions. This work was supported by Grants-in-Aid for Scientific Research from JSPS
(Nos.\ 16J05446 (J.O.), 15K17646 (H.T.) and 17H01110 (H.T.)) and the IBS under the project code IBS-R018-D1.

\end{acknowledgments}

\end{document}